\documentclass[fontset = windows,UTF8]{ctexart}

\usepackage{hyperref}
\usepackage{amsmath}
\usepackage{bm}
\usepackage{multirow}
\usepackage{verbatim}
\usepackage{float}
\usepackage{varwidth}

\usepackage{graphics}
\usepackage{epsfig}
\usepackage{pgf}
\usepackage{subfigure}
\usepackage{color}
\usepackage{pgfpages}
\usepackage{xcolor}
\usepackage{pifont}

\usepackage{amssymb}
\usepackage{amsfonts}
\usepackage{amsmath}

\usepackage[numbers,sort&compress]{natbib}

\usepackage{geometry}
\usepackage{cancel}
\usepackage{ulem}

\usepackage{comment}
\usepackage[format=hang,font=small,textfont=it]{caption}
\usepackage[nottoc]{tocbibind}
\usepackage[greek,english]{babel}
\usepackage{pstricks,pst-plot,pst-grad,pst-3dplot,pstricks-add,pst-node}

\usepackage{authblk}

\newcommand{\beq}[1]{\begin{eqnarray}\label{#1}}
	\newcommand{\eeq}{\end{eqnarray}}

\title{Entanglement entropy of particles in a perturbative scattering}
\author[1]{Jinbo Fan \thanks{jinbofan@outlook.com}}
\author[2]{Xuanting Ji \thanks{jixuanting@cau.edu.cn}} 
\author[1]{Xi-Jun Ren \thanks{997468504@qq.com}}
\affil[1]{School of Physics and Electronics, Henan University, Kaifeng, 475004, China}
\affil[2]{Department of Applied Physics, College of Science, China Agricultural University, Beijing 100083, China}

\begin{document}
	\maketitle         
\begin{abstract}
The entanglement among scattering particles in an exemplary quantum electrodynamics(QED) process is studied perturbatively.
To increase the computational accuracy, we need to consider virtual photon loop diagrams, which lead to infrared divergence. 
Therefore, when including higher order corrections in perturbative theory, we use the dressed state formalism proposed by Chuang and Faddeev-Kulish, which provides a finite S-matrix element. 
The entanglement entropy calculation shows that the entanglement is distributed linearly over the scattering cross section under the perturbation approximation.
\end{abstract}
	
\maketitle
\section{Introduction}
\label{intro}
The classical observables in scattering processes, such as the cross section and decay rate, are not sufficient to express the full information of scattering events. For example, the soft photons emitted during bremsstrahlung radiation carry a polarization vector, but their energy is too low to be detected, resulting in the loss of information about the scattering system. Entanglement, a characteristic quantity unique to quantum mechanics, has received extensive attention in the framework of quantum field theory \cite{Cardy:2004,Calabrese:2009,Horodecki:2009,Balasubramanian:2011wt,Hsu:2012gk}  and can be utilized to analyse infrared information in scattering processes \cite{Carney:2017jut,Gomez:2018war}. Furthermore, many quantum information concepts, such as entanglement entropy, error correction codes, and complexity, can improve our understanding of interaction field theory and spacetime geometry \cite{Takayanagi:2006,Pastawski:2015qua,Brown:2015lvg}.

Owing to the intriguing nature of entanglement, many authors have endeavoured to analyse the behaviours of entangled particles in the scattering process from various perspectives, identifying a linear relationship between entanglement entropy and the cross section of the process\cite{Park:2014hya,Peschanski:2016,Peschanski:2019yah,Fan:2017hcd,Fan:2017mth,Sampaio:2016,Semenoff:2016,Ratzel:2016qhg,Araujo:2019mni,Fedida:2022izl,Blasone:2024jzv}.
However, these studies focused primarily on the tree diagram model, 
simplifying the entire scattering process to two-body scattering. If we attempt to extend the above conclusion to a general perturbation case that incorporates higher order perturbation effects, we inevitably encounter infrared divergence caused by virtual photon loops. In the inclusive formalism, an additional scattering process, soft bremsstrahlung, is considered to eliminate this divergence. The cross sections of different processes are then summed to determine the contribution of the infrared-finite amplitude \cite{Weinberg:1965nx}.
This approach to counteract divergence is not performed at the level of the S-matrix, and involves only physical observables such as the cross section. In other words, when considering higher order contributions in perturbative theory, we cannot define a well-behaved S-matrix such that the amplitude of a given process without any soft photon emissions is finite \cite{Carney:2018ygh,Gomez:2017rau}. This complexity poses challenges for the analysis of the entanglement among scattered particles via the von Neumann entropy.

To address the above predicament, the formalism needs to satisfy two conditions: (i) the scattering process in QED is always accompanied by the emission of soft photons, and (ii) it retains a well-defined S-matrix for QED.	
These issues can be addressed via the dressed formalism proposed by Chung and Faddeev-Kulish \cite{Chung:1965zza,Kulish:1970ut}. In accordance with this description, the initial and final states are dressed with soft photon coherent states, which provide a finite $S$-matrix.
Recently, significant and potentially physically observable differences in how the two forms--the dressed formalism and the inclusive formalism--distribute quantum information during scattering have been identified\cite{Carney:2017jut,Carney:2018ygh,Carney:2017oxp}. In the inclusive formalism, the scattering final state is accompanied by many soft photons that are highly entangled with the remaining hard particles. The measurement of observable particles in the final state induces decoherence, and the scattering final state becomes a completely decohered mixed state. Consequently, describing infrared-finite quantum interference phenomena is impossible. In contrast, in the dressed formalism, the charged particles are enveloped by the soft photon cloud, and the infrared part is decoupled. Thus, the entire scattering process equates to an evolution from a pure state to another pure state, eliminating decoherence \cite{Semenoff:2019dqe}. In essence, the dressed formalism may enable a more fundamental description of the theory of QED scattering processes. Recently, many studies have extended this concept to gravitational scattering processes, particularly those involving the formation and evaporation of black holes, which may elucidate the black hole information paradox \cite{Choi:2017ylo,Choi:2018oel}.

The remainder of this paper is organized as follows. In Section II, we introduce the scattering process and address the entanglement between particles in the inclusive formalism. In Section III, the calculations are carried out via the dressed formalism. The conclusions are presented in Section IV.

\section{Entanglement entropy of particles in the inclusive formalism}
\label{sec:2}
We consider the scattering process of two fermionic fields in QED, $\Psi_A+\Psi_B\rightarrow \Psi_A+\Psi_B$, with the Hamiltonian $H=H_{\textup{free}}+H_{\textup{int}}$. For an elastic scattering process, the initial and final states can be described as a superposition of the basis of the free Hamiltonian, $H_{\textup{free}}=H_A\otimes H_B$,
\begin{align}
	\vert p,s;q,r\rangle=\sqrt{2E_{\textbf{p}}}a^{s\dagger}_{\textbf{p}}\vert0\rangle_A
	\otimes\sqrt{2E_{\textbf{q}}}b^{s\dagger}_{\textbf{q}}\vert0\rangle_B,
\end{align}
where $\textbf{p}$ and $\textbf{q}$ are 3-momenta, and s, r denote the spin of the fermions. The inner product between free particle states is defined as
\begin{align}
	\notag
	&\langle k,s^\prime;l, r^\prime\vert p,s;q,r\rangle
	\\ 
	&=2E_{\textbf{k}}2E_{\textbf{l}}
	(2\pi)^3\delta^{(3)}(\textbf{k}-\textbf{p})
	(2\pi)^3\delta^{(3)}(\textbf{l}-\textbf{q})
	\delta^{ss^\prime}\delta^{rr^\prime}.
\end{align}

\subsection{Entanglement Entropy}
\label{sec:21}
Consider a general initial state that is a superposition of spin eigenstates of two charged particles,
\begin{align}
	\vert\textup{ini}\rangle=\sum_{\sigma_a,\sigma_b}f_{\sigma_a\sigma_b}\vert p_a,\sigma_a;p_b\sigma_b\rangle,
\end{align}
where the coefficients satisfy $\sum_{\sigma_a,\sigma_b}\vert f_{\sigma_a\sigma_b}\vert^2=1$. The scattering final state is determined by the initial state and the $S$-matrix,
\begin{align}
	\notag
	\vert \textup{fin}\rangle=&\int \frac{d^3p_c}{(2\pi)^32E_{p_c}}\frac{d^3p_d}{(2\pi)^32E_{p_d}}
	\sum_{\sigma_c,\sigma_d}\vert p_c,\sigma_c;p_d,\sigma_d\rangle
	\\
&\times	\langle p_c,\sigma_c;p_d,\sigma_d\vert S\vert \textup{ini}\rangle.
\end{align}
The $T$ matrix is defined as
\begin{align}
	\notag
	&i\textbf{T}=\textbf{S}-1, 
	\\ \notag
	&\langle p_c,\sigma_c;p_d,\sigma_d\vert i\textbf{T} \vert p_a,\sigma_a;p_b,\sigma_b\rangle
	\\ \notag
	&=(2\pi)^4\delta^{(4)}(p_a+p_b-p_c-p_d)
	\times i\mathcal{M}(\sigma_a\sigma_b\rightarrow\sigma_c\sigma_d).
\end{align}

In the following, we describe the entanglement entropy between scattering particles. As discussed in \cite{Park:2014hya}, the entanglement distribution of the final state is contained in the density matrix, $\rho^f=\vert\textup{fin}\rangle\langle\textup{fin}\vert$. The reduced density matrix $\rho^f_A$ is obtained by tracing the degrees of freedom for part $B$, $\rho^f_A=\mathcal{N}^{-1}tr_B[\rho^f]$. 

When performing partial traces, the energy-momentum conversation factor, $(2\pi)^4\delta^{(4)}(p_i-p_f)$, must be considered. The entire scattering process is designed to occur in a large spacetime volume with duration $T$ and spatial volume $V$; these factors are artefacts caused by regulating the delta function \cite{Weinberg:1995}
\begin{align}
	\delta^{(3)}_V(\textbf{p}-\textbf{p}^\prime)&=\frac{V}{(2\pi)^3}\delta_{\textbf{p},\textbf{p}^\prime},
	\\ 
	\delta_T(E_i-E_f)&=\frac{1}{2\pi}\int_{-T/2}^{T/2}dt e^{i(E_i-E_f)t},
\end{align}
which implies that $V=(2\pi)^3\delta^{(3)}_V(0)$ and $T=(2\pi)\delta_T(0)$. The factors $T$ and $V$ are eliminated with proper normalization. To simplify the expressions, the following shorthand notations are used,
\begin{align}
	\notag
	&\mathcal{M}_{\sigma_c\sigma_d}\equiv\sum_{\sigma_a,\sigma_b}f_{\sigma_a\sigma_b}\mathcal{M}(\sigma_a\sigma_b\rightarrow\sigma_c\sigma_d),
	\\  \notag
	&\mathcal{F}_{\sigma_a\sigma_{a^\prime}}\equiv \sum_{\sigma_b}f_{\sigma_a\sigma_b}f^{\star}_{\sigma_{a^\prime}\sigma_b},\mathcal{A}_{\sigma_c\sigma_{c^\prime}}\equiv \sum_{\sigma_d}\mathcal{M}_{\sigma_c\sigma_d}\mathcal{M}^{\star}_{\sigma_{c^\prime}\sigma_d}.
\end{align}

Finally, the reduced density matrix $\rho^f_A$ can be written as
\begin{align}
	\notag
	\rho^f_A=&\frac{1}{\mathcal{N}}\biggr\{
	2E_{p_b}V2E_{p_a}V
	\begin{pmatrix}
		\mathcal{F}_{11}   &   \mathcal{F}_{12}   \\
		\mathcal{F}_{21}   &   \mathcal{F}_{22}   \\
	\end{pmatrix}  
	\otimes \frac{\vert p_a\rangle\langle p_a\vert}{2E_{p_a}V}
	\\ 
	&+\int d\Pi^{p_c}_2TV
	\begin{pmatrix}
		\mathcal{A}_{11}  &  \mathcal{A}_{12} \\
		\mathcal{A}_{21}  &  \mathcal{A}_{22} \\
	\end{pmatrix}
	\otimes \frac{\vert p_c\rangle\langle p_c\vert}{2E_{p_c}V}
	\biggr\},
\end{align}
where $\mathcal{N}$ is the normalization factor, which is fixed at  $\textup{tr}_A\rho^f_A=1$, and where $\int d\Pi^{p_c}_2$ is the relativistic invariant 2-body phase space,
\begin{align}
	\notag
	\int d\Pi^{p_c}_2\equiv&\int
	\frac{d^3p_c}{(2\pi)^32E_{p_c}}\frac{d^3p_d}{(2\pi)^32E_{p_d}}
	\\ \notag
	&\times(2\pi)^4\delta^{(4)}(p_a+p_b-p_c-p_d) .
\end{align}

To avoid divergence caused by collinearity, we analyse the entanglement of the final particles along a particular scattering angle $\Omega$ in momentum space. The reduced density matrix under the perturbation approximation is
\begin{align}
	\rho^f_{\Omega}(A)=
	\frac{T}{2E_{p_a}2E_{p_b}V}(\frac{d\Pi^{p_c}_2}{d\Omega})
	\begin{pmatrix}
		\mathcal{A}_{11}  &  \mathcal{A}_{12} \\
		\mathcal{A}_{21}  &  \mathcal{A}_{22} \\
	\end{pmatrix},
\end{align}	
has the eigenvalues $\{\alpha^2 r_i^A,i=1,2\}$, where $\alpha=e^2/4\pi$ is the fine-structure constant in QED.
Given the initial state, the relationship between the $S$-matrix elements and cross sections is
\begin{align}
	d\sigma_{\vert\textup{ini}\rangle}=\frac{1}{2E_a2E_b v_{rel}} d\Pi_2^{p_c}\vert\mathcal{M}(p_{\textup{ini}}\rightarrow p_c,p_d)\vert^2,
\end{align}
where $v_{rel}$ is the relative velocity between the incoming particles.
Thus, the trace of the density matrix $\rho^f_{\Omega}(A)$ is
\begin{align}
	\notag
	&Tr[\rho^f_\Omega(A)]
	\\ 
	&=\frac{T}{E_{p_a}E_{p_b}V}(\frac{d\Pi^{p_c}_2}{d\Omega})(\mathcal{A}_{11}+\mathcal{A}_{22})=T\Phi(\frac{d\sigma}{d\Omega})_{\vert\textup{ini}\rangle},
\end{align}
where $\Phi\equiv v_{rel}/V$ is the flux of the incoming particles.

The entanglement entropy between scattered particles along a particular momentum direction is
\begin{align}
	\notag
	S^f_{\Omega}(A)&=-\sum_i\alpha^2 r^A_i\log[\alpha^2 r_i^A]
	\\ 
	&=-\log[\alpha^2]T\Phi(\frac{d\sigma}{d\Omega})_{\vert\textup{ini}\rangle}+\mathcal{O}(\alpha^2).
\end{align}
The nondiagonal elements $\{\mathcal{A}_{12},\mathcal{A}_{21}\}$ appear in the $\alpha^2$ order and have negligible contributions under the perturbative approximation. By integrating over the entire momentum space, the leading order change in the entanglement entropy is found to be proportional to the cross section.

\subsection{IR divergence}
\label{sec:22}
\begin{figure}
	\centering
	\begin{varwidth}[htp]{\textwidth} 
		\vspace{0pt}
		\includegraphics[scale=0.25]{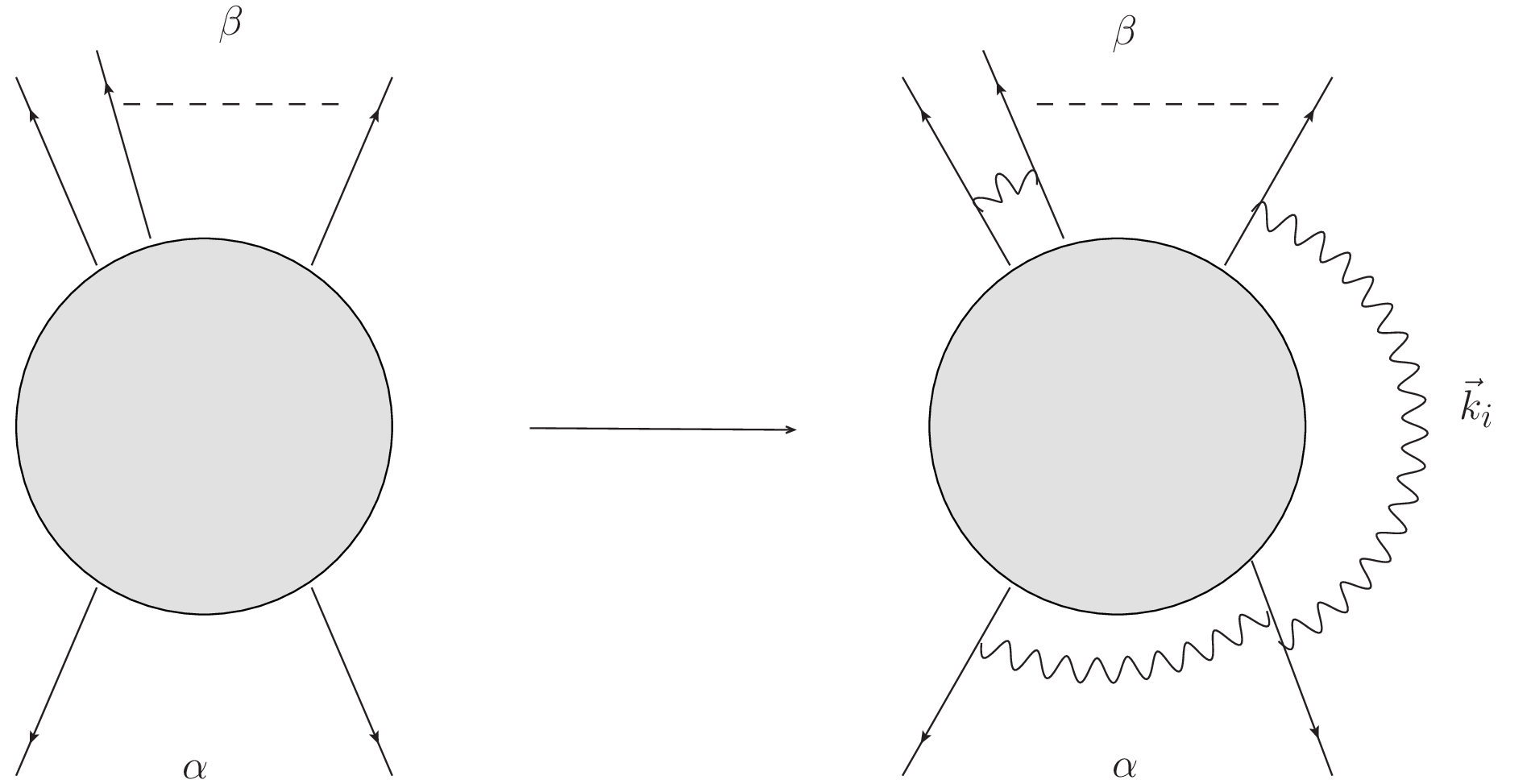}
	\end{varwidth}
	\caption{Higher order corrections are considered in an arbitrary process $\alpha\rightarrow\beta$ by introducing virtual photon loop diagrams. The straight lines are charged particles in the states $\vert\alpha\rangle$ and $\vert\beta\rangle$; the wavy lines are soft photons.}
	\label{fig:1}
\end{figure}

The above conclusion holds well under the tree level approximation. However, when considering higher order effects, we encounter infrared divergences caused by virtual photon loop diagrams, and the scattering with any emission of soft photons becomes zero.
As shown in Fig. $\ref{fig:1}$, for each virtual soft photon, the amplitude in the QED scattering process $\alpha\rightarrow\beta$ is of the form \cite{Weinberg:1965nx}
\begin{align}
	\notag
	&\mathcal{M}_{\beta\alpha}\longrightarrow\mathcal{M}_{\beta\alpha}
	\frac{1}{(2\pi)^4}\sum_{nm}e_ne_m\eta_n\eta_m(-ip_n\cdot p_m) 
	\int_{\lambda\le\vert \textup{k}_i\vert\le \Lambda}
	\\
	&\times\frac{d^4k_i}{[k_i^2-i\epsilon][p_n\cdot k_i-i\eta_n\epsilon][-p_m\cdot k_i-i\eta_m\epsilon]},
\end{align}
Here, $\lambda$ is the infrared cutoff scale, which is eventually set to zero, and $\Lambda$ is UV-cutoff for loop integration.
Moreover, $p_n$ and $e_n$ are the 4-momenta and charge of the $n$-th particle in the initial and final states, respectively, and $\eta_n$ is a sign factor with the value $+1$ for particles in the final state $\vert\beta\rangle$ and $-1$ for particles in the initial state $\vert\alpha\rangle$. 

The integral has a divergent imaginary part, but we do not explicitly write it, as it cancels out in the relevant density matrix.

In the case of a loop, the amplitude for instance
\begin{align}
	\label{1loop}
	\mathcal{M}_{\beta\alpha}^{(1,loop)}=\mathcal{M}_{\beta\alpha}[\frac{1}{2}B_{\beta\alpha}\ln(\frac{\lambda}{\Lambda})],
\end{align}
where $B_{\beta\alpha}$ denotes the positive kinematical factor for any nontrivial scattering. 
Obviously, the amplitude Eq. $(\ref{1loop})$ is divergent in the limit $\lambda\rightarrow0$.
One can consider the contribution of all the virtual photon loops, the amplitude
\begin{align}
	\label{loopamp}
	\mathcal{M}_{\beta\alpha}^\lambda=\mathcal{M}_{\beta\alpha}^\Lambda
	(\frac{\lambda}{\Lambda})^{B_{\beta\alpha}/2}.
\end{align}
When $\lambda\rightarrow 0$, virtual photon loops always contribute to making this amplitude vanish. $\mathcal{M}^{\Lambda}_{\beta\alpha}$ represents the scattering amplitude between undressed states, excluding the contribution of all virtual soft photons with energy below $\Lambda$.

\begin{figure}
	\centering
	\begin{varwidth}[htp]{\textwidth} 
		\vspace{0pt}
		\includegraphics[scale=0.18]{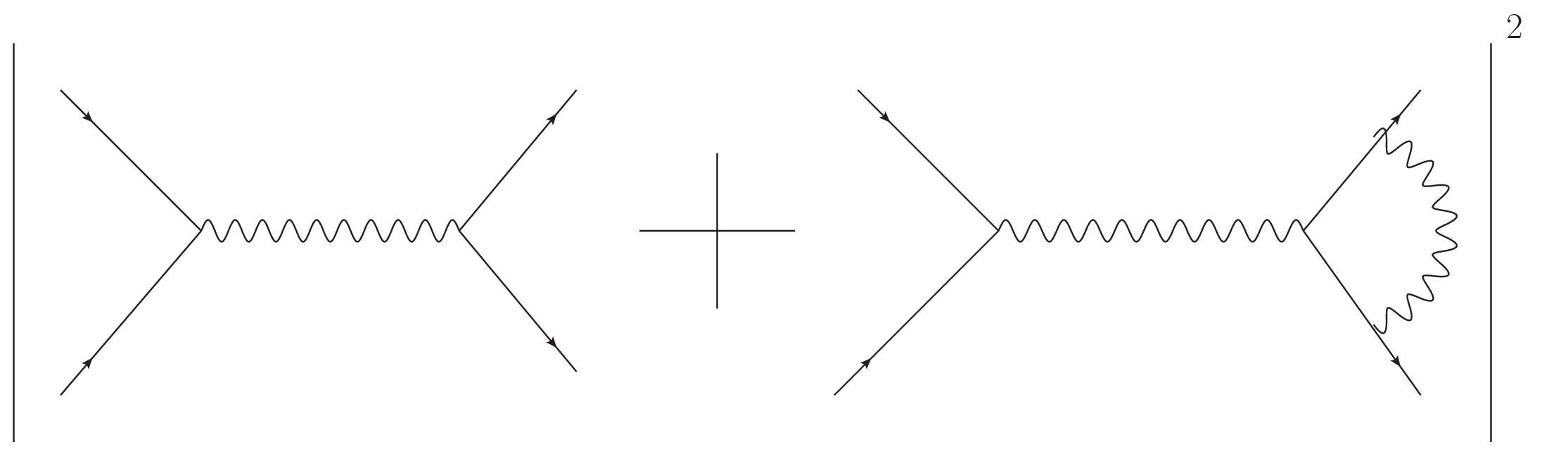}
	\end{varwidth}
	\qquad
	\qquad
	\begin{varwidth}[htp]{\textwidth}
		\vspace{0pt}
		\includegraphics[scale=0.18]{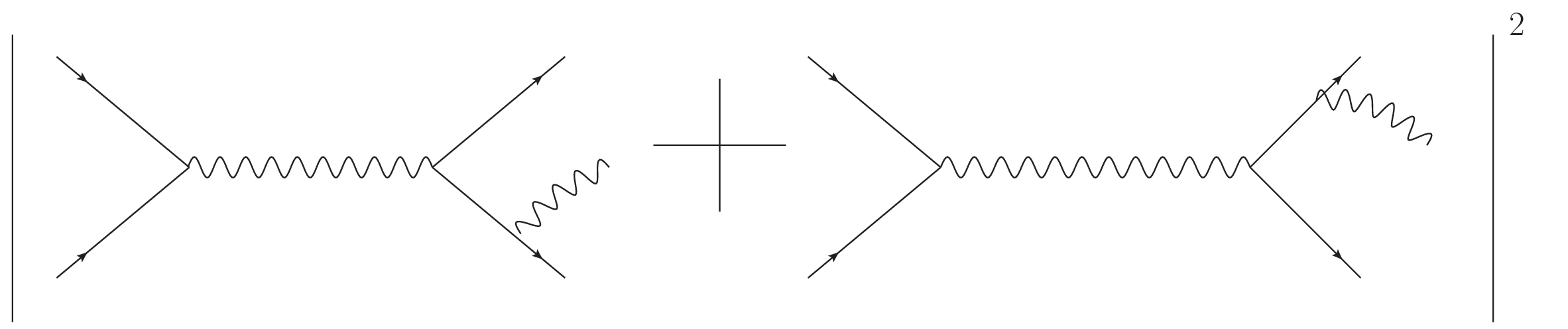}
	\end{varwidth}
	\caption{Left: The cross section for the $2\rightarrow2$ process $e^+e^-\rightarrow\mu^+\mu^-$ is IR divergent at order $e^6$;
		Right: The cross section for the $2\rightarrow3$ process $e^+e^-\rightarrow\mu^+\mu^-\gamma$ is IR divergent at order $e^6$. These two types of divergences can be cancelled out.}
			\label{fig:2}
\end{figure}

To address the infrared divergences caused by virtual soft photon loops, the traditional solution is to consider an additional indistinguishable process (bremsstrahlung) involving soft photons, which also introduces divergence. 
Using the QED scattering process $e^+e^-\rightarrow\mu^+\mu^-$ as an example (as shown in Fig. $\ref{fig:2}$), although the cross section for the process $e^+e^-\rightarrow\mu^+\mu^-$ is IR divergent at order $e^6$, as is the cross section for the related process $e^+e^-\rightarrow\mu^+\mu^-\gamma$, their sum is infrared finite,
\begin{align}
	\notag
	&\sigma_{inclusive}
	\\ \notag
	&=\sigma(e^+e^-\rightarrow\mu^+\mu^-)+\sigma(e^+e^-\rightarrow\mu^+\mu^-\gamma)
	=\textup{finite}.
\end{align}

At the level of the cross section, these two types of divergences can be cancelled out, thus preserving a finite transition probability.
Although the theoretical description is closely integrated with existing experimental approaches, a well-defined $S$-matrix is absent from the inclusive formalism because the amplitude of any scattering process involving finite soft photons is zero, complicating the analysis of the entanglement characteristics of the final state through the $S$-matrix. To address this issue, Chung and Faddeev-Kulish proposed dressed states in which charged particles are surrounded by a cloud of soft photons \cite{Chung:1965zza,Kulish:1970ut}. In this formalism, the S-matrix between any dressed state is finite, and the infrared divergence caused by soft photons is eliminated at the level of the scattering amplitude. Therefore, we can construct the final state from the $S$-matrix and calculate the von Neumann entropy of the final state.
Next, we use dressed states to analyse the entanglement of the observable charged particles in the scattering final state.

\section{Entanglement entropy of particles in dressed formalism}
\label{sec:3}
\subsection{Dressed state}
\label{sec:31}
In QED, purely hard scattering processes are forbidden by symmetry. Large gauge transformations require that hard asymptotic particles be accompanied by an infinite number of soft photons \cite{Kapec:2017tkm,Strominger:2017zoo}. In the dressed formalism, charged particles are viewed as systems surrounded by a cloud of soft photons, and the entire dressed state can be expressed as the product of a bare electron state and a coherent state of photons \cite{Gomez:2018war,Tomaras:2019sjq}	.
\begin{align}
	\notag
	&\Vert\alpha\rangle\!\rangle=\vert\alpha\rangle\otimes\vert D(\alpha)\rangle,
    \\
	&\vert D(\alpha)\rangle=\mathcal{N}_{\alpha}e^{\int_\lambda^{E_d}d^3\vec{k}\sum_l F_\alpha^{(l)}(\vec{k})a^{\dagger}_{l}(\textbf{k})}\vert 0\rangle,
\end{align}
where $a_{l}^{\dagger}(\textbf{k})$ creates photons with momentum $\textbf{k}$ and polarization vector $\epsilon_l^\mu(\textbf{k})$, and
\begin{align}
	F^{(l)}_{\alpha}(\textbf{k})=\sum_{n\in\alpha}\frac{e_n}{\sqrt{(2\pi)^32\vert\textbf{k}\vert}}\frac{p_n\cdot\epsilon^{\star}_{l}(\textbf{k})}{p_n\cdot k},
\end{align}
where the factor $\frac{\epsilon^{\star}_l(\textbf{k})}{\sqrt{(2\pi)^32\vert \textbf{k}\vert}}$ arises from the normalization of the photon wave function and where $e_n$ is the charge of the $n$th particle.
The normalization factor $\mathcal{N}_{\alpha}$ is given by
\begin{align}
	\mathcal{N}_\alpha=e^{-\frac{1}{2}\int_\lambda^{E_d} d^3\textbf{k}\sum_l F_{\alpha}^{(l)}(\textbf{k})F_{\alpha}^{\star(l)}(\textbf{k})}.
\end{align}
Similar to Eq. $(\ref{loopamp})$, this expression eventually gives an exponential function $(\frac{\lambda}{E_d})^{B_{\alpha}}$. Notably, the integral interval $(\lambda,E_d)$ indicates that only real soft photons with energies lower than the infrared energy $E_d$ are allowed, which are the photon clouds encapsulating the charged particles.

The scattering process $\alpha\rightarrow\beta$ is always accompanied by the emission of soft photons.
For the given initial state $\Vert\alpha\rangle\!\rangle$, the final state is determined by the well-defined $S$-matrix of QED,
\begin{align}
	\vert \textup{fin}\rangle=\sum_\beta\sum_{\gamma_n}\Vert\beta\gamma_n\rangle\!\rangle
	\langle\!\langle\beta\gamma_n\Vert S\Vert\alpha\rangle\!\rangle,
\end{align}
where $\Vert\beta\gamma_n\rangle\!\rangle\equiv\Vert\beta\rangle\!\rangle\otimes\vert\gamma_n\rangle$ represents the complete basis of dressed states, and where $\vert\gamma_n\rangle=a^{\dagger}(\textbf{k}_1)a^{\dagger}(\textbf{k}_2)\cdots a^{\dagger}(\textbf{k}_n)\vert0\rangle$ represents the emission of $n$ real soft photons.
From the leading soft theorems \cite{Weinberg:1965nx}, in the limit $k\rightarrow0$, the change in amplitude resulting from the emission of a soft photon to an arbitrary QED diagram is
\begin{align}
	\notag
	&\mathcal{M}_{\beta\alpha}\longrightarrow
	\mathcal{M}_{\beta\alpha}F^{(l_i)}_{\beta\alpha}(\textbf{k}_i),
	\\ 
	&F^{(l_i)}_{\beta\alpha}(\textbf{k}_i)=
	\frac{1}{\sqrt{(2\pi)^32\vert\textbf{k}_i\vert}}\sum_{n\in\alpha,\beta}\frac{e_n\eta_n(p_n\cdot\epsilon^{\star}_{l_i}(\textbf{k}_i))}{p_n\cdot k_i},
\end{align}
where $\eta_n=+1$ or $-1$ depending on whether it corresponds to the $\beta$ or $\alpha$ state. Thus, the matrix element $\tilde{S}_{\beta\gamma_n,\alpha}$ is evaluated between the dressed states,
\begin{align}
	\notag
	\tilde{S}_{\beta\gamma_n,\alpha}&=\langle\!\langle\beta\gamma_n\Vert S\Vert\alpha\rangle\!\rangle
	\\
	&=\langle\!\langle\beta\Vert S\Vert\alpha\rangle\!\rangle
	\Pi_i^n \sum_{l_i=1}^2F^{(l_i)}_{\beta\alpha}(\textbf{k}_i),
\end{align}
where $\langle\!\langle\beta\Vert S \Vert\alpha\rangle\!\rangle=\tilde{S}_{\beta\alpha}$ is
the $S$-matrix between the incoming and outgoing dressed states.

According to the computations in \cite{Tomaras:2019sjq,Gomez:2018war}, the dressed amplitude and the undressed amplitude are related as
\begin{align}
	\tilde{\mathcal{M}}_{\beta\alpha}=(\frac{E_d}{\lambda})^{B_{\beta\alpha}/2}\mathcal{M}_{\beta\alpha}^\lambda.
\end{align}
Combined with Eq. $(\ref{loopamp})$, $\mathcal{M}_{\beta\alpha}^\lambda=\mathcal{M}_{\beta\alpha}^\Lambda
(\frac{\lambda}{\Lambda})^{B_{\beta\alpha}/2}$, we obtain
\begin{align}
	\label{Smatrix}
	\tilde{\mathcal{M}}_{\beta\alpha}=(\frac{E_d}{\Lambda})^{B_{\beta\alpha}/2}
	\mathcal{M}^{\Lambda}_{\beta\alpha}.
\end{align}
Clearly, this quantity is infrared finite; moreover, the dependence on the scale $\Lambda$ is cancelled between the exponential factor and  $\mathcal{M}^{\Lambda}$.

Thus, the final state takes the following superposition form
\begin{align}
	\notag
	\vert \textup{fin}\rangle=&\Vert \alpha\rangle\!\rangle+i\sum_\beta
	(2\pi)^4\delta^{(4)}(p_\alpha-p_\beta)
	\tilde{\mathcal{M}}_{\beta\alpha}\Vert\beta\rangle\!\rangle
	\\
	&\otimes e^{\int_{E_d}^\epsilon d^3\textbf{k}	\sum_lF_{\beta\alpha}^{(l)}(\textbf{k})a^{\dagger}_l(\textbf{k})}\vert0\rangle,
\end{align}
where the part $e^{\int_r^\epsilon d^3\textbf{k}	\sum_lF_{\beta\alpha}^{(l)}(\textbf{k})a^{\dagger}_l(\textbf{k})}\vert0\rangle$ corresponds to the radiation state $\vert\gamma(\alpha,\beta)\rangle$ in \cite{Gomez:2018war}. In the combined formalism, the energy scales are as follows: $\lambda<E_d<\epsilon<\Lambda$. These energy scales categorize soft photons in the scattering process into dressing and radiation photons. The first type exists in both the initial and final states, where charged particles are dressed. The dressing depends solely on their respective state and consists of soft photons with energies in the interval $\{\lambda,E_d\}$.
The second type exists only in the final state, comprising soft photons with energies in the interval $\{E_d,\epsilon\}$, which are involved in the scattering interaction. This implies that the radiation state $\vert\gamma(\alpha,\beta)\rangle$ depends on both the initial and final states of the charged particles. Obviously, in the case where $E_d=\epsilon$, we obtain the dressed formalism in which there is no radiation.

\subsection{Hard density matrix}
\label{sec:32}

\begin{figure}
	\centering
	\begin{varwidth}[htp]{\textwidth} 
		\vspace{0pt}
		\includegraphics[scale=0.25]{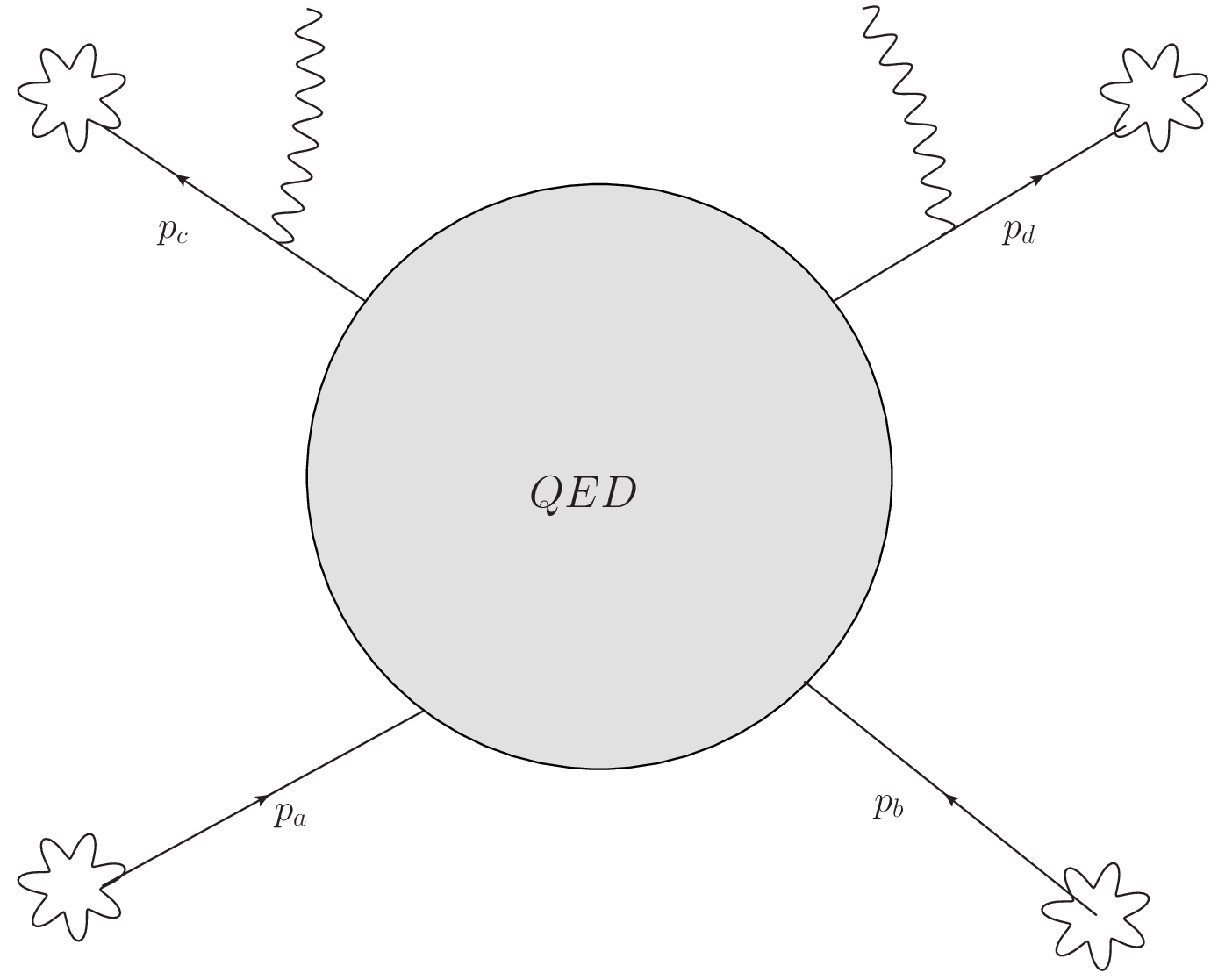}
	\end{varwidth}
	\caption{In the dressing formalism, the scattering process $\alpha\rightarrow\beta$ is accompanied by two types of soft photons, $\alpha=\{e_a,p_a,\sigma_a;e_b,p_b,\sigma_b\}$ and $\beta=\{e_c,p_c,\sigma_c;e_d,p_d,\sigma_d\}$. The dressed photons exist in the initial $\vert\alpha\rangle$ and final states $\vert\beta\rangle$, surrounding the charged particles. The radiation photons exist only in the final state $\vert\beta\rangle$.}
		\label{fig:3}
\end{figure}

Our main task is to analyse the entanglement of the observable charged particles in the final state.
As shown in Fig. $\ref{fig:3}$, we expand the discussion in Sect. $\ref{sec:2}$ by employing dressed states to describe the scattering process of charged particles, $\Psi_A+\Psi_B\rightarrow\Psi_A+\Psi_B$, which incorporates higher order corrections via loop diagrams.
We start with a general dressed initial state,
\begin{align}
	\Vert\textup{ini}\rangle\!\rangle=\sum_{\sigma_a,\sigma_b}\tilde{f}_{\sigma_a\sigma_b}\Vert p_a,\sigma_a;p_b,\sigma_b\rangle\!\rangle,
\end{align}
where the coefficients satisfy $\sum{\sigma_a,\sigma_b}\vert \tilde{f}_{\sigma_a\sigma_b}\vert^2=1$. 
The final state is given by the action of a finite $S$-matrix on the initial dressed state
\begin{align}
	\notag
	\Vert \textup{fin}\rangle\!\rangle=&\Vert \textup{ini}\rangle\!\rangle+i\int d\Pi_2^{p_c}
	\tilde{\mathcal{M}}_{\sigma_c\sigma_d}\Vert p_c,\sigma_c;p_d,\sigma_d\rangle\!\rangle
	\\
	&\otimes e^{\int_{E_d}^{\epsilon}d^3\textbf{k}
		\sum_{l}F_{\beta\alpha}^{(l)}(\textbf{k})a^{\dagger}_l(\textbf{k})}\vert0\rangle,
\end{align}
where $\tilde{\mathcal{M}}_{\sigma_c,\sigma_d}\equiv\sum_{\sigma_a,\sigma_b}\tilde{f}_{\sigma_a\sigma_b}\tilde{\mathcal{M}}(\sigma_a\sigma_b\rightarrow\sigma_c\sigma_d)$. 

According to the sensitivity of the detector and the chosen energy cutoff $\epsilon$, we separate the asymptotic Hilbert spaces into soft and hard factors $\mathcal{H}=\mathcal{H}_H\times\mathcal{H}_S$. In this section, we focus on the entanglement of the hard degree of freedom between the scattering particles. Hard entanglement is measured by mutual information,
\begin{align}
	I(A_h,B_h)=S(A_h)+S(B_h)-S(A_h,B_h),
\end{align}
where $S(A_h)$, $S(B_h)$ and $S(A_hB_h)$ are the von Neumann entropy of the hard part in the scattering final state.

Using the dressed final state $\beta^\prime=\{e_{c^\prime},p_{c^\prime},\sigma_{c^\prime};e_{d^\prime},p_{d^\prime},\sigma_{d^\prime}\}$ and tracing over all soft degrees of freedom in the final state, we obtain the reduced density matrix of the hard part,
\begin{align}
	\notag
	&\rho^f_{(A_hB_h)}
	\\ \notag
	=&\frac{1}{\tilde{\mathcal{N}}}\biggr\{
	\sum_{\sigma_a\sigma_b;\sigma_{a^\prime}\sigma_{b^\prime}}
	f_{\sigma_a\sigma_b}f^{\star}_{\sigma_{a^\prime}\sigma_{b^\prime}}
	\vert p_a,\sigma_a;p_b,\sigma_b\rangle\langle p_a,\sigma_{a^\prime};p_b,\sigma_{b^\prime}\vert
	\\ \notag
	&-\sum_{\sigma_c,\sigma_d;\sigma_{c^\prime},\sigma_{d^\prime}}
	\int d\Pi_2^{p_c}
	\int d\Pi_2^{p_{c^\prime}}
	(\frac{\lambda}{E_d})^{\frac{1}{2}B_{\beta\beta^\prime}}
	(\frac{\epsilon}{E_d})^{A_{\beta\beta^\prime,\alpha}}
	\\
	&\times\mathcal{\tilde{M}}_{\sigma_c\sigma_d}
	\mathcal{\tilde{M}}^{\star}_{\sigma_{c^\prime}\sigma_{d^\prime}}
	\vert p_c,\sigma_c;p_d,\sigma_d\rangle\langle p_{c^\prime},\sigma_{c^\prime};p_{d^\prime},\sigma_{d^\prime}\vert\biggr\},
\end{align}
where $\tilde{\mathcal{N}}$ is the normalization factor, which is fixed at  $\textup{tr}_{A_hB_h}[\rho^f(A_hB_h)]=1$. 

(i) The first IR factor $(\frac{\lambda}{E_d})^{\frac{1}{2}B_{\beta\beta^\prime}}$ arises from tracing over dressing photons,
\begin{align}
	\notag
	&(\frac{\lambda}{E_d})^{\frac{1}{2}B_{\beta\beta^\prime}}
	\\ 
	&=Exp[-\frac{1}{2}\int^{E_d}_\lambda\frac{d^3\textbf{k}}{(2\pi)^32\vert\textbf{k}\vert}
	\sum_{n,n^\prime\in\beta,\beta^\prime}\frac{\eta_n\eta_{n^\prime}e_ne_{n^\prime}(p_n\cdot p_{n^\prime})}{(p_n\cdot k)(p_{n^\prime}\cdot k)}],
\end{align}
and
\begin{align}
	B_{\beta\beta^\prime}&=-\frac{1}{8\pi^2}\sum_{n,n^\prime\in\beta,\beta^\prime}\frac{e_ne_{n^\prime}\eta_n\eta_{n^\prime}}{\beta_{nn^\prime}}\ln[\frac{1+\beta_{nn^\prime}}{1-\beta_{nn^\prime}}],
	\\
	&\beta_{nn^\prime}=\sqrt{1-\frac{m_n^2m_{n^\prime}^2}{(p_n\cdot p_{n^\prime})^2}}.
\end{align}
Notably, $\eta_n=+1$ if $n\in\beta$, whereas $\eta_n=-1$ if $n\in\beta^\prime$.
Clearly, if $0<\beta_{nn^\prime}<1$, we have $\beta_{nn^\prime}=0$ if and only if $p_c=p_{c^\prime}, p_d=p_{d^\prime}$. Combining the energy-momentum conservation factor $\delta^{(4)}(p_a+p_b-p_c-p_d)$ with the kinetic energy factor $(\frac{\lambda}{E_d})^{B_{\beta\beta^\prime}}$, the density matrix elements along the momentum direction $\{p_c,p_{c^\prime}\}\vert_{p_c\neq p_{c^\prime}}$ vanish in the limit $\lambda\rightarrow0$.

(ii) The second IR factor $(\frac{\epsilon}{E_d})^{A_{\beta\beta^\prime,\alpha}}$ arises from the contribution of soft radiation photons,
\begin{align}
	A_{\beta\beta^\prime,\alpha}=-\frac{1}{8\pi^2}\sum_{m\in\alpha,\beta; m^\prime\in\alpha,\beta^\prime}\frac{e_me_{m^\prime}\eta_m\eta_{m^\prime}}{\beta_{mm^\prime}}\ln[\frac{1+\beta_{mm^\prime}}{1-\beta_{mm^\prime}}],
\end{align}
The factor $\eta_m=-1$ if $n\in\alpha$, and $\eta_m=+1$ if $n\in\{\beta,\beta^\prime\}$. 
Note that the energy conservation condition is disregarded here; thus, such soft radiation photons should satisfy $\sum_r\omega_r\le \Lambda$.
When the restriction condition $\theta(\Lambda-\sum_r\omega_r)$ is considered, the IR radiation changes to
\begin{align}
	(\frac{\epsilon}{E_d})^{A_{\beta\beta^\prime,\alpha}}\rightarrow
	(\frac{\epsilon}{E_d})^{A_{\beta\beta^\prime,\alpha}}\times
	\mathcal{F}(\frac{\epsilon}{\Lambda},A_{\beta\beta^\prime,\alpha}).
\end{align}
For $\epsilon/\Lambda=\mathcal{O}(1)$ and $A_{\beta\beta^\prime,\alpha}<<1$, the factor $\mathcal{F}(\frac{\epsilon}{\Lambda},A_{\beta\beta^\prime,\alpha})$ is close to unity.

When the energy scale is $E_d=\epsilon$, the theory reverts to the dressed formalism.
For simplicity, we define 
$\mathcal{\tilde{A}}_{\sigma_c\sigma_d\sigma_{c^\prime}\sigma_{d^\prime}}= \mathcal{\tilde{M}}_{\sigma_c\sigma_d}\mathcal{\tilde{M}}^{\star}_{\sigma_{c^\prime}\sigma_{d^\prime}} $, the hard reduced density matrix along the momentum direction $\{\vec{p}_c=\vec{p}_{c^\prime},\vec{p}_d=\vec{p}_{d^\prime}\}$,
\begin{align}
	\notag
&\rho^f_{\Omega}(A_hB_h)=
\\
&\frac{T}{2E_{p_a}2E_{p_b}V}(\frac{d\Pi^{p_c}_2}{d\Omega})
\begin{pmatrix}
	\mathcal{\tilde{A}}_{1111}   &   \mathcal{\tilde{A}}_{1112} &   \mathcal{\tilde{A}}_{1121} &   \mathcal{\tilde{A}}_{1122}   \\
	\mathcal{\tilde{A}}_{1211}   &   \mathcal{\tilde{A}}_{1212}  &   \mathcal{\tilde{A}}_{1221} &   \mathcal{\tilde{A}}_{1222}  \\
	\mathcal{\tilde{A}}_{2111}   &   \mathcal{\tilde{A}}_{2112} &   \mathcal{\tilde{A}}_{2121} &   \mathcal{\tilde{A}}_{2122}   \\
	\mathcal{\tilde{A}}_{2211}   &   \mathcal{\tilde{A}}_{2212}  &   \mathcal{\tilde{A}}_{2221} &   \mathcal{\tilde{A}}_{2222}  \\	
\end{pmatrix},
\end{align}	
has eigenvalues $\alpha^2r_i^H\equiv\alpha^2 r_i^{(0)}+\alpha^3 r_i^{(1)}+\cdots+\alpha^{2+N}r_i^{(N)},i=\{1,2,3,4\}$ and $\alpha^{2+N}r_i^{(N)}$ arising from the contribution of N-loop diagrams.
According to Eq. ($\ref{Smatrix}$), which describes the dressed $S$-matrix, we choose the detector scale $\epsilon=\Lambda$. The dressed amplitude has $\tilde{\mathcal{M}}(\sigma_a\sigma_b\rightarrow\sigma_c\sigma_d)=
\mathcal{M}^{\Lambda}$. 
Thus, the trace of the density matrix is
\begin{align}
	\notag
	Tr[\rho^f_\Omega(A_hB_h)]
	&=\frac{T}{2E_{p_a}2E_{p_b}V}(\frac{d\Pi^{p_c}_2}{d\Omega})\vert\mathcal{M}^\Lambda\vert^2
	\\
	&=T\Phi(\frac{d\sigma}{d\Omega})_{\textup{incl}},
\end{align}
where $\mathcal{\tilde{A}}_{1111}+\mathcal{\tilde{A}}_{1212}+\mathcal{\tilde{A}}_{2121}+\mathcal{\tilde{A}}_{2222}=\vert\mathcal{\tilde{M}}\vert^2$. $(\frac{d\sigma}{d\Omega})_{\textup{incl}}$ is the differential cross section in the inclusive formalism, with the cutoff for the virtual photon loop integral set at $\Lambda$.

Finally, the leading order of the entanglement entropy between soft and hard particles in the momentum direction is
\begin{align}
	\notag
	S^f_\Omega(A_hB_h)&=-\sum_i\alpha^2 r_i^H\log[\alpha^2 r^H_i]
	\\
	&=-\log[\alpha^2]T\Phi(\frac{d\sigma}{d\Omega})_{\textup{incl}}+\mathcal{O}(\alpha^2).
\end{align}
The physical quantity obtained by integrating the scattering angles across the entire momentum space describes the entanglement between the hard particles and the soft radiation in the final state, which can be viewed as a response to the infrared information of the system. In a real scattering experiment, we cannot measure both the charged particles and the surrounding soft photon clouds; instead, we can measure only the charged particles. Therefore, we need a theory that can describe the phenomenon of quantum interference in terms of the final particles, and the scattering process described by this theory does not exhibit full decoherence. This theoretical realization can be obtained in the dressed state, as the hard particles in the final state and the soft radiation are entangled. Moreover, the corresponding hard density matrix is infrared finite and does not lead to full decoherence, which is also responsible for well-measured phenomena such as radiation damping.

Moreover, there is further discussion regarding the role of soft radiation in the scattering process. In the dressed formalism, the cross section for the emission of soft photons with energies less than $E_d$ is suppressed \cite{Choi:2019rlz}. When the real photon energy $\omega_\gamma>E_d$, the entangled entropy (per unit time, per particle flux) between the hard and soft particles is proportional to the inclusive cross section of the given initial state \cite{Irakleous:2021ggq,Toumbas:2023qbo}.

\subsection{Mutual information}
\label{sec:33}

The reduced density matrix of the hard parts in $A$ and $B$ is represented by $\rho_{A_hB_h}$. By tracing over the hard degree of freedom in $B$, we obtain the reduced density matrix for the hard part in $A$, $\rho^f_{A_h}$.
To simplify the expressions, the following shorthand notation is used in this section,
\begin{align}
	\notag
	\mathcal{\tilde{A}}^c_{\sigma_c\sigma_{c^\prime}}\equiv \sum_{\sigma_d}\mathcal{\tilde{M}}_{\sigma_c\sigma_d}\mathcal{\tilde{M}}^{\star}_{\sigma_{c^\prime}\sigma_d},
	~~
	\mathcal{\tilde{A}}^d_{\sigma_d\sigma_{d^\prime}}\equiv \sum_{\sigma_c}\mathcal{\tilde{M}}_{\sigma_c\sigma_d}\mathcal{\tilde{M}}^{\star}_{\sigma_c\sigma_{d^\prime}}.
\end{align}
For a particular scattering angle $\Omega$ in momentum space, the reduced density matrix,
\[	
\rho^f_{\Omega}(A_h)=
\frac{T}{2E_{p_a}2E_{p_b}V}(\frac{d\Pi^{p_c}_2}{d\Omega})
\begin{pmatrix}
	\mathcal{\tilde{A}}^c_{11}  &  \mathcal{\tilde{A}}^c_{12} \\
	\mathcal{\tilde{A}}^c_{21}  &  \mathcal{\tilde{A}}^c_{22} \\
\end{pmatrix},
\]
has eigenvalues $\alpha^2 r_i^{c}\equiv \alpha^2 r_{c,i}^{(0)}+\alpha^3 r_{c,i}^{(2)}+\cdots+\alpha^{2+N} r_{c,i}^{(N)}, i=1,2$, where $\alpha^{2+N}r_{c,i}^{(N)}$ comes from the contribution of the N-loop diagram. Using the relation $\mathcal{\tilde{A}}^c_{11}+\mathcal{\tilde{A}}^c_{22}=\mathcal{\tilde{A}}_{1111}+\mathcal{\tilde{A}}_{1212}+\mathcal{\tilde{A}}_{2121}+\mathcal{\tilde{A}}_{2222}$, we find that the trace of the matrix is
\begin{align}
	Tr[\rho^f_{\Omega}(A_h)]
	=\frac{T}{2E_{p_a}2E_{p_b}V}(\frac{d\Pi^{p_c}_2}{d\Omega})\vert\mathcal{M}^{\Lambda}\vert^2.
\end{align}
Thus, the entanglement entropy of the $A$ hard part at the leading order is
\begin{align}
	S^f_\Omega(A_h)
	=-\log[\alpha^2]T\Phi(\frac{d\sigma}{d\Omega})_{\textup{incl}}+\mathcal{O}(\alpha^2).
\end{align}

Similarly, given that the amplitude in momentum space satisfies the relation $\mathcal{M}(\theta)=\mathcal{M}(\pi-\theta)$, the leading order of the entanglement entropy of the hard part in B is
\begin{align}
	S^f_\Omega(B_h)&=-\log[\alpha^2]T\Phi(\frac{d\sigma}{d\Omega})_{\textup{incl}}+\mathcal{O}(\alpha^2).
\end{align}

Finally, the entanglement entropy between the hard parts in $A$ and $B$ along the scattered direction is
\begin{align}
	I_{\Omega}(A_h,B_h)&=S_\Omega(A_h)+S_\Omega(B_h)-S_\Omega(A_hB_h)
	\\ \notag
	&=-\log[\alpha^2]T\Phi(\frac{d\sigma}{d\Omega})_{\textup{incl}}+\mathcal{O}(\alpha^2).
\end{align}
where $(\frac{d\sigma}{d\Omega})_{\textup{incl}}$ is the differential cross section with virtual photons loop integral cut off at $\Lambda$.

The presence of soft photons in the dressed state does not affect the degree of entanglement between scattered charged particles. When considering higher order perturbative computations, the noninfrared virtual photons contribute to the entanglement between charged particles, thus causing a change in the entanglement entropy in the momentum direction that remains proportional to the differential cross section of the process. Within perturbation theory, the leading order change in the entanglement entropy between scattered particles is generally proportional to the cross section.

\section{Conclusion}
\label{sec:4}

In this work, we analyse the behaviour of entangled particles during QED scattering according to perturbative theory. When only the contributions from the tree level diagrams are considered, the final scattering state can be constructed using the initial state and the $S$-matrix, and the entanglement of the scattering particles can be analysed via von Neumann entropy. The leading order change in the entanglement entropy between the scattered particles is proportional to the cross section. To extend this conclusion to a broader context, higher order contributions to the scattering amplitude in perturbative computations must be considered. However, the virtual photon loops in the Feynman diagrams induce infrared divergence, and the scattering amplitude with any soft photon emission is zero. Consequently, it becomes impossible to treat the final state as a pure state and utilize its von Neumann entropy to analyse the entanglement between the scattered particles.

Therefore, we adopted a different method to address infrared divergence: the dressed formalism proposed by Chung and Faddeev-Kulish. In this formalism, the initial and final states are both dressed with coherent states of soft radiation, and the $S$-matrix elements between these dressed states take finite values. Taking the perturbative processes in QED scattering as an example, we analyse the entanglement properties of dressed particles. Although there are different descriptions for QED processes, the influence of the scattering process on the entanglement between particles is reflected primarily in the cross section, and the leading order change in the entanglement entropy between charged particles is found to be proportional to the cross section.

In recent years, many researchers have analysed the relationships between classical observables in the gravitational framework and quantum observables in field theory, such as entanglement entropy, mutual information, and quantum teleportation. According to the holographic principle, gravitational theory and quantum field theory should be considered together, with spacetime emerging from entanglement. To further explore the relationship between the entanglement entropy and observable measurements in scattering theory, this research should be extended to include strongly coupled models. Within the AdS=CFT correspondence, both the scattering amplitude and entanglement entropy in strongly coupled field theory are linked to minimal surfaces in bulk gravity theory. Moreover, the holographic understanding of the relationship between scattering and entanglement entropy could be explored through the AdS=CFT correspondence.

\section*{Acknowledgment}

The authors would like to thank Jin-Xuan Zhao for useful discussions. This
work is supported by the National Natural Science Foundation of China (Grants No. 11947046, No. 12105075 and No. 12405078).

\section*{Data Availability Statement}
The authors confirm that there are no associated data in the paper.

% BibTeX users please use
% \bibliographystyle{}
% \bibliography{}

\begin{thebibliography}{}
\bibitem{Cardy:2004}
P. Calabrese, J. L. Cardy, Entanglement entropy and quantum field theory, J. Stat. Mech. \textbf{0406}, P06002 (2004).
\bibitem{Calabrese:2009}
P. Calabrese, J. L. Cardy, Entanglement entropy and conformal field theory, J. Phys. A. \textbf{42}, 504007 (2009).

\bibitem{Horodecki:2009}
R. Horodecki, P. Horodecki, M. Horodecki  and K. Horodecki, Quantum entanglement, Rev. Mod. Phys. \textbf{81}, 865 (2009).

\bibitem{Balasubramanian:2011wt}
V. Balasubramanian, M. B. McDermott and M. Van Raamsdonk, Momentum-space entanglement and renormalization in quantum field theory, Phys. Rev. D \textbf{86}, 045014 (2012).

\bibitem{Hsu:2012gk}
T. C. L. Hsu, M. B. McDermott and M. Van Raamsdonk, Momentum-space entanglement for interacting fermions at finite density, JHEP \textbf{1311}, 121 (2013).

\bibitem{Carney:2017jut}
D. Carney, L. Chaurette, D. Neuenfeld and G. Semenoff, Infrared quantum information, Phys. Rev. Lett. \textbf{119}(18), 180502 (2017).

\bibitem{Gomez:2018war}
C. Gómez, R. Letschka and S. Zell, The Scales of the Infrared, JHEP \textbf{09}, 115 (2018).

\bibitem{Takayanagi:2006} S. Ryu, T. Takayanagi, Holographic Derivation of Entanglement Entropy from AdS/CFT, Phys. Rev. Lett. \textbf{96}, 181602 (2006).

\bibitem{Pastawski:2015qua}
F. Pastawski, B. Yoshida, D. Harlow and J. Preskill, Holographic quantum error-correcting codes: Toy models for the bulk/boundary correspondence, JHEP \textbf{1506}, 149 (2015).

\bibitem{Brown:2015lvg} 
A. R. Brown, D. A. Roberts, L. Susskind, B. Swingle and Y. Zhao, Complexity, action, and black holes, Phys. Rev. D \textbf{93}, 086006 (2016).

\bibitem{Park:2014hya}
S. Seki, I. Y. Park and S. J. Sin, Variation of Entanglement Entropy in Scattering Process, Phys. Lett. B \textbf{743}, 147 (2015).

\bibitem{Peschanski:2016}
R. Peschanski, S. Seki, Entanglement Entropy of Scattering Particles, Phys. Lett. B. \textbf{758}, 89 (2016).

\bibitem{Peschanski:2019yah}
R. Reschanski, S. Seki, Evaluation of Entanglement Entropy in High Energy Elastic Scattering, Phys. Rev. D \textbf{100}(7), 076012 (2019).

\bibitem{Fan:2017hcd}
J. Fan, Y. Deng and Y. C. Huang, Variation of entanglement entropy and mutual information in fermion-fermion scattering, Phys. Rev. D \textbf{95}, 065017 (2017).

\bibitem{Fan:2017mth} 
J. Fan and X. Li, Relativistic effect of entanglement in fermion-fermion scattering,
Phys. Rev. D \textbf{97}, 016011 (2018).

\bibitem{Sampaio:2016} 
R. Faleiro, A.S. Helder, R. Pavão, H. Alexander, B. Hiller, H. Blin and M. Sampaio,  Perturbative approach to entanglement generation in QFT using the S matrix, J. Phys. A \textbf{53}, 365301 (2020).


\bibitem{Semenoff:2016} 
D. Carney, L. Chaurette, G. Semenoff, Scattering with partial information, arXiv:1606.03103[hep-th].

\bibitem{Ratzel:2016qhg}
D. Rätzel, M. Wilkens and R. Menzel, Effect of polarization entanglement in photon-photon scattering, Phys. Rev. A \textbf{95}, 012101 (2017).

\bibitem{Araujo:2019mni} 
J. B. Araujo, B. Hiller, I. G. da Paz, M. M. Ferreira, Jr., M. Sampaio and H. A. S. Costa, Measuring QED cross sections via entanglement, Phys. Rev. D \textbf{100}, 105018 (2019).

\bibitem{Fedida:2022izl}  
S. Fedida and A. Serafini, Tree-level entanglement in quantum electrodynamics, Phys. Rev. D \textbf{107},116007 (2023).

\bibitem{Blasone:2024jzv}
M. Blasone, S. De Siena, G. Lambiase, C. Matrella and B. Micciola, Complete complementarity relations in tree level QED processes, Phys. Rev. D \textbf{111}, 016007 (2025).

\bibitem{Weinberg:1965nx}
S. Weinberg, Infrared photons and gravitons, Phys. Rev. \textbf{140}, B516-B524 (1965).

\bibitem{Carney:2018ygh}
D. Carney, L. Chaurette, D. Neuenfeld and G. Semenoff, On the need for soft dressing, JHEP \textbf{09}, 121 (2018).

\bibitem{Gomez:2017rau}
C. Gómez, R. Letschka and S. Zell, Infrared Divergences and Quantum Coherence, Eur. Phys. J. C \textbf{78}(8), 610 (2018).

\bibitem{Chung:1965zza}
V. Chung, Infrared Divergence in Quantum Electrodynamics, Phys.Rev. \textbf{140}, B1110-B1122 (1965).

\bibitem{Kulish:1970ut}
P.P. Kulish, L.D. Faddeev, Asymptotic conditions and infrared divergences in quantum electrodynamics, Theor. Math. Phys. \textbf{4} 745 (1970), Teor.Mat.Fiz. \textbf{4} 153-170 (1970).

\bibitem{Carney:2017oxp}
D. Carney, L. Chaurette, D. Neuenfeld and G. Semenoff, Dressed infrared quantum information, Phys. Rev. D \textbf{97}(2), 025007 (2018).

\bibitem{Semenoff:2019dqe}
G. Semenoff, Entanglement and the Infrared, Springer Proc.Math.Stat. \textbf{335} 151-166 (2019).

\bibitem{Choi:2017ylo}
S. Choi, R. Akhoury, BMS Supertranslation Symmetry Implies Faddeev-Kulish Amplitudes, JHEP \textbf{02}, 171 (2018).

\bibitem{Choi:2018oel}
S. Choi, R. Akhoury, Soft Photon Hair on Schwarzschild Horizon from a Wilson Line Perspective, JHEP \textbf{12}, 074 (2018).

\bibitem{Weinberg:1995}
S. Weinberg, \textit{The Quantum Theory of Fields I}, Cambridge University Press, N. Y. (1995).

\bibitem{Kapec:2017tkm}
D. Kapec, M. Perry, A. P. Raclariu and A. Strominger, Infrared Divergences in QED, Revisited, Phys. Rev. D \textbf{96}(8), 085002 (2017).

\bibitem{Strominger:2017zoo}
A. Strominger, Lectures on the Infrared Structure of Gravity and Gauge Theory, arXiv:1703.05448[het-th].

\bibitem{Tomaras:2019sjq}
T. Tomaras and N. Toumbas, IR dynamics and entanglement entropy, Phys. Rev. D \textbf{101} 6, 065006 (2020).

\bibitem{Choi:2019rlz}
S. Choi, R. Akhoury, Subleading soft dressings of asymptotic states in QED and perturbative quantum gravity, JHEP \textbf{09}, 031 (2019).

\bibitem{Irakleous:2021ggq}
A. Irakleous, T. Tomaras and N. Toumbas, Soft photon radiation and entanglement, Eur. Phys. J. C \textbf{81}(8), 745 (2021).

\bibitem{Toumbas:2023qbo}
N. Toumbas and A. Irakleous, Scattering, IR dynamics and entanglement, PoS CORFU2022, 152 (2023).


\end{thebibliography}

\end{document}